\documentstyle[11pt,aaspp4]{article}  

\lefthead{Cavaliere, Menci \& Tozzi}
\righthead {Groups and Clusters of Galaxies}

\begin{document}
\title{Hot Gas in Clusters of Galaxies: the Punctuated Equilibria Model}
\author{A. Cavaliere}
\affil{Astrofisica, Dipartimento Fisica, II Universit\`a di Roma,\\
via Ricerca Scientifica 1, 00133 Roma, Italy}
\author{N. Menci}
\affil{Osservatorio Astronomico di Roma,
 via Osservatorio, 00040 Monteporzio, Italy}
\and
\author{P. Tozzi
}
\affil{ Astrofisica, Dipartimento Fisica, II Universit\`a di Roma,\\
via Ricerca Scientifica 1, 00133 Roma, Italy \\
 Present address: Department  of Physics and Astronomy, The Johns Hopkins
 University, Baltimore, MD 21218, USA.
}
\begin{abstract}
We develop our  model of  ``punctuated equilibria'' for the hot 
intra-cluster 
gas emitting powerful X-rays. The model considers 
the gravitational potential wells set by the dark matter as they evolve 
by  hierarchical clustering and engulf outer gas; it assumes that
 the gas re-adjusts to a new hydrostatic equilibrium after each merging event. 
Before merging the gas is heated at the virial temperature 
when bound in subclusters; at early $z$ it is 
preheated by supernova activity following star formation. 

In detail, we compute analytically the following steps: 
the dynamic histories of dark matter halos with their 
merging events; the associated infall of gas into a halo, 
with compressions and shocks estabilishing 
the conditions at the cluster boundary; the updated 
disposition of the gas in the potential well matching such conditions; 
the statistical convolution of observable quantities over the merging 
histories. 

 For the individual objects from groups to clusters, the model 
 yields profiles of density and surface brightness 
 with no free parameters; in particular, the so-called $\beta$ 
parameter is itself an outcome of the model, and the polytropic index $\gamma$ 
is internally constrained to a narrow range.
We obtain declining temperature profiles, 
 and profiles for the density and for the surface brightness 
 shallower in groups compared with clusters; our model groups also contain 
a lower baryonic fraction on average, but with a scatter considerably 
larger. 

We present various key quantities over the whole range from groups to clusters.
In particular, we predict in different cosmologies the statistical 
correlation $L-T$ of luminosity with temperature; 
similarly, we derive the correlation $R_X-T$ for 
the size of the X-ray emitting region.  The intrinsic scatter in both 
correlations is also predicted. 
 \end{abstract}
{\bf key words:} galaxies: clusters: general -- intergalactic 
medium -- X-rays: galaxies

\newpage
\section{INTRODUCTION}
Groups and clusters of galaxies constitute large, nearly virialized 
condensations. Within their virial radii $R$ ranging from $1/2$ to about 2 Mpc 
the density contrasts attain or exceed $\delta\rho/\rho_u\sim 2\,10^2$ 
relative to the background, and the 
corresponding masses  $M$ 
range from some $10^{13}$ to $10^{15}$ $M_{\odot}$,  mainly in dark matter (DM).

These structures contain a large baryonic fraction $f\simeq 0.1$ 
in the form of a hot intra-cluster plasma (ICP) at 
 temperatures
$k\,T\sim GM\,m_H/10 R\sim 0.5 - 10$ keV ($m_H$ is the proton mass) 
 with particle densities up to 
$n\sim 10^{-3}$ cm$^{-3}$. The ICP emits by thin thermal bremsstrahlung 
 X-ray luminosities $L\propto n^2\,T^{1/2}\,R^3$ ranging 
from $10^{42}$ to $10^{45}$ erg/s. 

Recent observations of individual objects  resolved in energy and angle  
 with the ASCA and with the SAX satellites 
 indicate radial temperature profiles declining outwards 
(Markevitch, Sarazin \& Henriksen 1997; 
Molendi, private communication; see also Hughes, Gorenstein \& Fabricant 1988).  
In several cases the thermal structure is 
complicated by hot spots (Honda et al. 1997; Markevitch et al. 1998). 
 As for the statistical aspects, a steep correlation  close to the overall 
form $L\propto T^3$ 
is known to hold for local clusters,   
but with a substantial scatter  (Edge \& Stewart 1991;  Mushotzky 1994). 
Recently the observations 
have sampled higher redshifts out to $z\sim 0.5$ 
(Tsuru et al. 1996; Mushotzky \& Scharf 1997) finding little 
significant evolution. Low-temperature, local systems 
have been also sampled, finding there indications of a slope steeper 
yet (Ponman et al. 1996). For $kT> 5$ keV a 
 flattening toward $L\propto T^2$ has been detected by Allen \& Fabian (1997). 

In the near future the AXAF mission will substantially 
improve the space-resolved 
spectroscopy of many individual clusters, and shortly after the mission 
XMM will greatly enlarge the statistics of the $L-T$ correlation. 
Corresponding upgrades are called for in the theoretical 
understanding of the complex astrophysics concerning 
both the DM and the ICP over 
the full range from groups to clusters.  

The force approach uses numerical computations for both the DM and the 
ICP, striving for wide dynamic range and complete 
hydrodynamics. 
The N-body simulations have shown, in accord 
with the hierarchical clustering picture (see Peebles 1993),  
 the evolution of the DM halos 
to occur largely through a sequence of merging and accretion events 
which involve generally smaller partners down to nearly diffuse 
matter (see, e.g., Tormen, Bouchet \& White 1997), and are correlated with 
  the surrounding large scale structures (Colberg et al. 1998)

As for the ICP, pioneering work by Schindler \& M\"uller (1993) 
 taken up by Roettiger, Stone \& Mushotzky (1998)  used 
3D Eulerian codes with adaptive mesh and advanced 
shock-capturing methods to study how the large 
merging events of the DM halos affect the ICP component. 
 The outcomes show how such events produce anisotropic shocks and 
 non-uniform compressions, resulting in 
a complex thermal structure lasting a few Gyrs. 

At the other extreme, a sequence of 
radial, highly resolved Eulerian computations  
(progressing from Perrenod 1980 to Takizawa \& Mineshige 1998a) have shown that 
isotropic accretion of smooth gas also causes a strong and 
slowly expanding shock, which 
remains close to the (growing) virial radius for some dynamical times.  

Most recently, state-of-the-art N-body codes coupled with advanced hydro
(Bryan \& Norman 1998, Gheller et al. 1998) have been 
run on supercomputers, aiming at 
 resolutions below 100 kpc in rich clusters as necessary for deriving reliable 
luminosities. But such simulations 
are hard-pressed in implementing at the same time 
the full physics of the ICP. In fact, preheating at temperatures 
$\sim 0.5$ keV is expected from stellar formation and evolution to 
 supernovae  (see Renzini 1997, and 
references therein); this is particularly relevant for the shallower potentials 
of groups where $T$ is close to $0.5$ keV. Inclusion of the preheating  
in the numerical work is technically taxing, as it involves 
cooling, star formation and energy feedbacks resolved down to 
subgalactic scales; but in its absence 
 the simulations produce a correlation of the form $L\propto T^2$ 
at all temperatures, at variance with the data. 
Suginohara \& Ostriker (1998) stress how delicate may become at high 
resolutions the balance of cooling and feedbacks, and how difficult becomes 
reproducing ICP cores as observed. 
Thus for now and for some time to come it will be hard to combine 
 into a realistic numerical picture 
wide dynamic range from galaxies to large scale structures, stellar preheating 
and large statistics. 

The state of the numerical approach and the challenge from the data 
 motivates us to present here 
an {\it analytic} model which includes, though in a 
simplified form, the physical processes outlined above. 
We describe the cluster history as a sequence of {\it punctuated equilibria} 
(PE). That is to say, we envisage such history as a sequence of hierarchical 
merging episodes of the DM halos which we compute analytically (with its 
variance) in 
the framework of the standard hierarchical clustering, specifically 
using the so-called ``extended Press \&  Schechter theory'' (Bond et al. 1991; 
 Bower 1991; Lacey \& Cole 1993). We stress that 
such episodes cause in the gas shocks of various strengths 
depending on the mass ratio of the merging subclusters, ranging from 
 nearly adiabatic compressions for comparable clumps up to shocks with 
high Mach numbers in the accretion of loose gas. Our point is that the most
effective such shocks and compressions overlap to 
provide the boundary conditions for the  new hydrostatic equilibrium 
 to which the ICP is assumed to re-adjust. 

The PE model as presented here takes up our previous work 
(Cavaliere, Menci \& Tozzi 1997, 1998; hereafter CMT97, CMT98), but differs 
 in that the histories of the DM halos are now computed 
{\it analytically} rather than based on Monte Carlo simulations. 
This goes beyond the technical aspect since it 
allows us to explore efficiently the dependences of the density,  
temperature and luminosity on the parameters of the 
clusters and on the cosmology. 
In the same vein, the present approach allows us to quantify the 
 connection between slope and scatter of the $L-T$ correlation 
and the cosmological scenario. 
In addition, the parameters of the ICP thermal state are 
 now fixed or bounded in terms of constraints internal to the model, and 
{\it new} predictions are presented.  

In Sect. 2 we describe the PE model and our computational steps: 
 in Sect. 2.1 we recall the 
statistical formalism for hierarchically merging DM halos; 
in Sect. 2.2, we derive from 
the mass ratios involved in each merging episode 
the boundary conditions for the plasma equilibrium; 
 in Sect. 2.3 we compute from such boundary conditions 
the ICP equilibrium;  in Sect. 2.4 we derive 
the statistics of $L$ and of the size $R_X$ using the formalism of Sect. 2.1.
 In Sect. 3 we give, and compare with the observations, the model results 
 for the profiles $n(r)$ and $T(r)$, for the surface brightness 
$\Sigma (r)$, for 
 the relations $M-T$ and $f-T$, and for the correlations 
$R_X-T$ and $L-T$. 
The final Sect. 4 is devoted to discussion and conclusions. 
\section{THE PUNCTUATED EQUILIBRIA MODEL}
The X-ray bolometric luminosity of a cluster is given in its basic 
dependences by 
\begin{equation}
L\propto \int_o^{r_2}\,n^2(r)\,T^{1/2}(r)\,d^3r~.  
\end{equation}
Here $T(r)$ is temperature in the plasma and $r_2$
is the cluster boundary, that we take to be close to 
the virial radius $R \propto M^{1/3}\,\rho^{-1/3}$, where
$\rho (z)\propto (1+z)^3$ is the DM density in the cluster, 
 proportional to the average cosmic DM density $\rho_u(z)$ at formation.  

  It will be convenient to separate 
the internal profiles $n(r)$ and $T(r)$ from 
 their boundary conditions at $r_2$. As to the latter, 
 the infalling gas is  expected to become supersonic near $r_2$ 
(see, e.g., Perrenod 1980; Takizawa \& Mineshige 1997) so
 that a shock front will 
form there. The conservations across the shock of mass, 
energy and stresses yield the Rankine-Hugoniot conditions, i.e., 
the temperature and density jumps from the outer 
values $T_1$ and $n_1$ to $T_2$ and $n_2$ just interior to $r_2$ 
(spelled out in Sect. 2.2). 
Then the luminosity may be rewritten in the form 
\begin{equation}
L\propto r_2^3\;n_2^2\,T_2^{1/2}\;
\int_0^1 d^3x\,\Big[{n(x)\over n_2}\Big]^2\,
\Big[{T(x)\over T_2}\Big]^{1/2}~, 
\end{equation}
 where $x\equiv r/r_2$.

Note that the 
values $n_2$ and $T_2$ at the boundary are not uniquely determined 
 by the cluster mass $M$; rather, they are related to the outer
 values 
$n_1$ and $T_1$ by the named shock conditions. In turn, 
 $n_1$ is fixed by $n_1 \propto f_u \, \rho_u /m_H$, 
in terms of the universal 
baryonic fraction $f_u$; whereas $T_1$ is determined  only  statistically, 
 through the diverse merging histories ending up in the mass $M$. 
 Specifically, as explained in detail in Sects. 2.1 and 2.2, in 
each merging episode $T_1$ takes on the values appropriate to 
 the  other merging partner, constituted by 
subclumps or even by smooth gas. 
In sum, a given dark mass $M$ admits a set of ICP equilibrium states 
 characterized by different boundary conditions, each corresponding to a 
different realization of the dynamical merging history. 
 It is the convolution over such set which provides the 
average values of $L$ and $R_X$, and their scatter. 

So the development of our PE model 
proceeds along the following steps: \hfill\break
i) we first give the 
statistics of the current DM halo of mass $M$ and of the merging clumps 
 $\Delta M$; 
\hfill\break
ii) we compute the shock strength relating  at the boundary the 
 inner values $T_2$ and $n_2$ to the exterior ones 
$T_1$ and $n_1$, as a function of $M$ and 
$\Delta M$; 
\hfill\break
iii) from such boundary values, we compute the internal 
profiles $T(x)/T_2$ and $n(x)/n_2$ 
for the post-merging hydrostatic equilibrium, 
 involving the  cluster potential and hence $M$; 
\hfill\break
iv) we convolve the results of steps ii) and iii) with the statistics i). 

Below we describe  these steps in turn. 
\subsection{Histories of the Dark Matter Halos}
 Here we recall the basic merging probabilities 
provided by the ``extended Press \& Schechter theory'', 
see Lacey \& Cole (1993; 1994). This is based on  
the dark halos formed by hierarchical merging of smaller structures.
 
The halo mass distribution at the cosmic time $t$ 
 is given by the standard Press \& Schechter (1974) formula 
\begin{equation}
N(m,t)=\sqrt{{2\over \pi}}\;{\delta_{c}(t)\,\rho_u\over \,M^2_{o} }\;
\Big|{{d\ln \sigma}\over {d\ln m}}\Big|\; {m^{-2}\over {\sigma(m)}}\;
e^{-{\delta_{c}(t)^2\over {2\,\sigma^2(m)}}}~, 
\end{equation}
where the masses $m\equiv M/M_o$ are normalized to the current value 
$M_{o}=0.6\, 10^{15}\,\Omega_0\,h^{-1}~M_{\odot}$ (i.e., to the mass 
enclosed within a sphere of 8$\,h^{-1}$ Mpc), and 
$\delta_{c}(t)=\delta_{co}\,D(t)$ is a critical threshold for the 
collapse and virialization of the primordial density perturbations. 
The local value $\delta_{co}$ 
 depends weakly on the cosmological parameters, while the growth factor 
$D(t)$ sensitively depends on them. 
The mass variance $\sigma(m)$ is computed 
in terms of the perturbation spectrum; 
for definitness, we use the CDM spectra given 
and discussed by  White, Viana \& Liddle (1996). 
For $\Omega=1$ we adopt the primordial 
 ``tilted'' index $n_p=0.8$; for $\Omega_o<1$ we adopt $n_p=1$. 
The associated  normalizations are taken from the COBE/DMR data 
(Gorski et al. 1998), and expressed in terms of  
the amplitude $\sigma_8$ 
at the relevant scale of $8\,h^{-1}$ Mpc 
(see Bunn \& White 1997). 

Corresponding to eq. (3), 
 the probability distribution 
that a given mass $m$ at time $t_o$ has a progenitor of mass 
$m'$ at time $t_1<t_o$ reads 
\begin{equation}
{d f\over dm'}(m',t_1|m,t_o)={\delta_{c}(t_1)-\delta_{co}\over
 (2\pi)^{1/2}(\sigma'^2-\sigma^2)^{3/2} }\,{m\over m'}\,
\big|{d\sigma'^2\over dm'}\big|\; exp{ 
\Bigg\{-
 {[\delta_{c}(t_1)-\delta_{co}]^2\over 2\,(\sigma'^2-\sigma^2) }
\Bigg\}
}~,
\end{equation}
where  $\sigma'$ is the mass variance at the scale $m'$. 
On the other hand, at a given time $t$ a progenitor $m'$ increases its mass 
by a merging event with a clump of mass $\Delta m$ 
(producing a cluster with mass $m=m'+\Delta m$), with 
the probability distribution per unit time given by 
$${d^2 p(m'\rightarrow m'+\Delta m)\over d\Delta m\,dt} = 
\Big({2\over \pi}\Big)^{1/2}\,
\Big|{d\,ln(\delta_c)\over dt}\Big|
\,\Big|{d\,ln \sigma \over dm}(m'+\Delta m)\Big|
\,{\delta_c(t)\over \sigma(m'+\Delta m)}\times
$$
\begin{equation}
\times {1\over \Big[ 1-\sigma^2(m'+\Delta m)/\sigma^2(m') \Big]^{3/2} }
\; exp{
\Bigg\{- 
{\delta_c^2(t)\over 2}\,
\Big[ {1\over \sigma^2(m'+\Delta m)} - {1\over \sigma^2(m')} \Big] 
\Bigg\}~.
}
\end{equation}
We have compared the analytical probabilities above with the results 
of the Monte Carlo code developed  by P. Tozzi to simulate 
the hierarchical merging history of halos, based on 
the excursion set approach of Bond et al. (1991). 
 A realization from the Monte Carlo simulations is shown in fig. 1a as an 
illustration of the basic process of DM halo growth. 
To show the agreement of the two approaches for a relevant quantity,  
 we plot in fig. 1b the probability distribution 
of progenitors of mass $m'$ at different $z$ 
 that end up in a given mass $m$ at $z=0$, 
computed  from the Monte Carlo and from the eqs. above. In fig. 1c 
we show as a function of $m$ 
 the fraction of objects which, during the last 2 Gyrs, accreted mass 
in events involving comparable clumps
(specifically, those with mass ratio 1/2.5), and in events involving very unequal 
 clumps (with mass ratio 1/10); during that interval, 
more than 60 \% of the clusters with $M\geq 10^{15}\,M_{\odot}$ will 
have merged with clumps 
 smaller than $M/10$. 
\subsection{Boundary Conditions }
The pre-shock temperature in a merging event 
is that of the infalling gas. If the latter is contained in a 
 sufficiently deep potential well, $T_1$ is the virial 
 temperature $T_{1v}\propto \Delta m/r$ 
of the secondary merging partner; on using 
$r\propto (\Delta m/\rho)^{1/3}$ this writes 
\begin{equation}
k\,T_{1v}=4.5\,(\Delta m)^{2/3}\,(\rho/\rho_o)^{1/3}~{\rm keV}, 
\end{equation}
where the numerical 
coefficient is taken from Hjorth, Oukbir \& van Kampen (1998). 
Where necessary, the $z$-dependence 
of $\rho/\rho_o=(1+z)^3$ is converted to $t$-dependence following 
the standard FRW cosmologies. 

But an independent lower bound 
$kT_{1*}\approx 0.5$ keV is provided by 
preheating of diffuse external gas, due to 
feedback energy inputs following star formation and evolution 
all the way to supernovae (David et al. 1995; 
Renzini 1997). We recall that 
 preheating temperatures in excess of 
$ 0.1$ keV are believed to constitute 
essential complements to the hierarchical clustering  
picture to prevent the ``cooling catastrophe'' 
from occurring, see White \& Rees (1978); 
Blanchard, Valls Gabaud \& Mamon (1992). In point of fact, Henriksen \& 
White (1996) find from X-rays evidence for diffuse gas at $0.5 - 1$ keV 
 in the outer regions of a number of clusters.  
So in the following the actual value of $T_1$ will be 
\begin{equation}
T_1=max\,\big[T_{1v},T_{1*}\big]~. 
\end{equation} 

Given $T_1$, the boundary conditions 
for the ICP in the cluster 
is  set by the strength of the shocks separating the inner 
from the infalling gas. 
 We report here from CMT98 the explicit expression
of the post-shock temperature $T_2$ for three degrees of freedom and
for a nearly hydrostatic post-shock condition with $v_2<< v_1$, 
assuming the shock velocity to match the growth rate of the 
virial radius R(t):
\begin{equation}
kT_2={{\mu m_H v_1^2}\over 3}\Big[ {{(1+\sqrt{1+\epsilon})^2}\over 4}
+ {7\over{10}}\epsilon -{{3}\over {20}}{{\epsilon^2}\over{(1+\sqrt{1+
\epsilon})^2}}\Big]\, . 
\end{equation}
Here $\epsilon\equiv 15 kT_1/4 \mu m_H v_1^2$ and $\mu$ is the average 
molecular weight;  the inflow velocity $v_1$ is set 
by the potential drop across the region of nearly free fall, to read 
$v_1 \simeq \sqrt{-\phi_2/m_H}$ in terms of the potential $\phi_2$ at $r_2$.  
For a ``cold inflow'' with 
$\epsilon <<1$ the shock is {\it strong}, and the expression simplifies to 
$kT_2\simeq {{\mu m_H v_1^2}/3} + 3kT_1/2$. 
Instead, for $\epsilon 
\gtrsim 1$
the shock is {\it weak}, and $T_2\simeq T_1$ is recovered as expected.  
Note that  $T_2$ depends through both $T_1$ and $v_1^2$ 
on the mass $\Delta m$  of the clump being accreted. 

>From $T_2$ and $T_1$, 
the density jump at the boundary $n_2/n_1$ is found to read 
(see CMT97)
\begin{equation}
{n_2\over n_1} = 
2\,\Big(1-{T_1\over T_2}\Big)+\Big[4\, 
\Big(1-{T_1\over T_2}\Big)^2 + {T_1\over T_2}\Big]^{1/2}~.
\end{equation}
It is seen that the density jump takes on the limiting value $n_2/n_1=4$ 
for very strong shocks, while the adiabatic approximation 
$n_2/n_1\approx 1+3(T_2-T_1)/2T_1$ is recovered for weak shocks.  
\subsection{Hydrostatic Equilibrium }
We adopt the polytropic temperature description 
$T(x)/T_2= [n(x)/n_2]^{\gamma-1}$, with the 
index $\gamma$ in the range $1\leq \gamma\leq 5/3$ to begin with. 
 In terms of the virial temperature $T_v$ (see Sarazin 1988), 
eq. (2) writes 
$L\propto r_2^3\,(n_1/\rho)^2\,(n_2/ n_1)^2\,T_v^{1/2}\,\rho^2
\big(T_2/ T_v\big)^{1/2}\,
\overline{[n(r)/n_2\big]^{2+(\gamma-1)/2}}$, 
where the bar denotes the integration 
 over the emitting volume $r^3\leq r_2^3$, and $\rho$ is the average 
DM density in the cluster, proportional 
 to $\rho_u$ and so to $n_1$ (see under eq. 2). 
 The radius $r_2$ may be rewritten in terms of the 
  temperature $T_v\propto m/r_2\propto \rho\,r_2^2$. 
 We finally obtain 
\begin{equation}
L\propto \Big({n_2\over n_1}\Big)^2\,T_v^2\,\rho^{1/2}\,
\Bigg[{T_2\over T_v}\Bigg]^{1/2}\,
\overline{[n(r)/n_2\big]^{2+(\gamma-1)/2}} ~.
\end{equation}

The underlying assumption is that after a merging event 
 the cluster re-adjusts to a hydrostatic equilibrium 
 with boundary conditions $n_2$, $T_2$ corresponding to 
 its dynamical history (see eq. 6-9). 
Actually, this requires sound propagation times 
 shorter than the dynamical timescale taken anyway by the DM to relax
to its own steady configuration; the condition is seen to be satisfied, 
 though marginally, for all merging events except for the {\it rare} ones 
 involving comparable clumps. 
As we discuss in detail in the concluding \S 4, the observations and the 
hydrodynamical N-body simulations concur in supporting 
not only the hydrostatic equilibrium approximation for the relevant merging 
events, but also its parametrization with a 
polytropic equation of state. 

The ratio $n(x)/n_2$ is obtained 
from the hydrostatic equilibrium 
$dP/m_H\,n\,dr=-G\,M(<r)/r^2=-d\phi/dr$ with the  polytropic pressure 
$P(r)= kT_2\,n_2\,\big[{n(r)/n_2}\big]^{\gamma}$.  This yields 
(see Cavaliere \& Fusco Femiano 1978; Sarazin 1988, and bibliography therein) 
the profiles 
\begin{equation}
{n(r)\over n_2}=\Big[{T(r)\over T_2}\Big]^{1/(\gamma-1)}=
\Big\{1+{\gamma-1\over \gamma}\,\beta\,
\big[\tilde{\phi}_2-\tilde{\phi}(r)\big]\Big\}^{1/(\gamma-1)}~,
\end{equation}
where $\tilde{\phi}\equiv \phi/\mu\,m_H\,\sigma_2^2$ is the potential
normalized  to the associated one-dimensional DM velocity dispersion at $r_2$. 
The ICP disposition in eq. (11) relative to the DM depends on the parameter 
\begin{equation}
\beta = \mu m_H \sigma_2/kT_2~,
\end{equation}
 and is further modulated by the second parameter $\gamma$, to 
yield as the latter increases flatter profiles $n(r)$ and steeper $T(r)$. 
In our PE 
$\beta$ is by $T_2$ given by eq. (8), and considering the statistics 
 of $T_2$ we obtain the results discussed in \S 3.1; the other parameter 
 $\gamma$ will be bounded as also discussed there. 

We shall focus on the ``universal'' forms of $\phi(r)$ and $\sigma(r)$ given
by Navarro, Frenk \& White (1997). 
 When relevant, we will discuss also results for the (simplified) 
 King potential (see Sarazin 1988; see also Adami et 
al. 1998) 
where the DM itself has a core, and somewhat fatter ICP cores obtain. 
We shall also discuss the steeper cusp 
 found by Moore et al. (1997) in  highly resolved, CDM simulations; 
correspondigly, we still obtain a core-like ICP distribution, 
albeit slightly slimmer.
Actually, in hydrostatic equilibrium a DM cusp flatter than 
$\rho(r) \propto r^{-2}$ -- corresponding to a gravitational force flatter 
then $r^{-1}$ -- 
implies at the centre a finite ICP density $n_c$ but a high derivative, which
however  at observable resolutions is flattened by a modest increase of 
$\gamma$. 

\subsection{Statistics}
Our purpose is to compute the average value of $L$ and its dispersion,  
 associated with a given cluster mass $m$. We re-iterate from Sect. 2 that 
 the diverse merging histories ending up in such a mass give rise to 
 a set of equilibrium states characterized by different values of 
$T_2$, $n_2$. These are 
   related by eqs. (8) and (9) to the values of $T_1$ associated with the 
 the clump of mass $\Delta m$ incoming onto a cluster progenitor. 
 So to meet our purpose we must sum over the shocks produced at a time $t'<t$ 
in all possible progenitors $m'$ (weighting with their number) 
by the accreted clumps $\Delta m$ (weighting with their merging rate); 
finally, we integrate over times $t'$ from an effective lower limit 
$t-\Delta t$.   

The average $L$ is then given by 
\begin{equation}
\langle L\rangle = Q\,\int_{t-\Delta t}^t\,dt'
\,\int_0^m\,dm'\,\int_0^{m-m'}\,d\Delta m
\,{df\over dm'}(m',t'|m,t)\,
{d^2 p(m'\rightarrow m'+\Delta m)\over d\Delta m\,dt'}
\,L~; 
\end{equation}
and the variance is given by 
\begin{equation}
\langle \Delta L^2\rangle = Q\,\int_{t-\Delta t}^t\,dt'
\int_0^m\,dm'\,\int_0^{m-m'}\,d\Delta m\,
\,{df\over dm'}(m',t'|m,t)\,
{d^2 p(m'\rightarrow m'+\Delta m)\over d\Delta m\,dt'}
\Big(L-\langle L\rangle\Big)^2~. 
\end{equation}
Higher moments -- if needed in case of non-Poissonian statistics --  are 
 given by similar expressions; 
the full distribution of $L$ requires aimed computations or simulations, as 
 noted by CMT98. 
In the integrals, the luminosity  $L\, \big[T_2(m'),T_1(\Delta m')\big]$ 
depends on $m'$ and $\Delta m'$ through the boundary conditions discussed
 in Sect. 2.2. 
The compounded probability distribution in eqs. (13) and (14) has been 
normalized to 1 (we do not write down 
the normalization factor $Q$ for the sake of simplicity). 
The effective lower limit for the integration over masses is set 
 as follows. 

The merging events relevant to $\langle L\rangle$  and to 
$\langle \Delta L^2\rangle$ after eqs. (13) and (14) are those lasting enough 
 as to overlap with similar ones. 
Since $\Delta t \propto \Delta m/v_1\,\rho\,r^2$, the above condition results in an 
effective lower limit for the masses entering eq. (13) and (14); 
physically, lumps with masses smaller than such limit yield a small
 mass flux and so produce shocks which 
 dissipate before new clumps income. We pinned down the minimum $\Delta m$ 
 numerically, by looking at the saturation of $\langle L\rangle$ 
for increasing values of $\Delta t$. This occurs at $\Delta t\approx 0.7\,t_d$ 
 which corresponds to a lower limit about $m/20$ for $\Delta m$ on using 
$r^2\propto \Delta m^{2/3}$. 

Note that the above procedure acts like an effective mass weight. 
Heuristically, this may be seen with the $\Delta m$ and $t$ 
integrations interchanged; then the lower limit contains $\Delta t (\Delta m)$
 which must be convolved with the distribution of 
$\Delta m$.  But for $\Delta t>0.7\,t_d$ the resulting average saturates; 
so we have written the $t$-integration as the outmost integral, 
having adopted the lower limit for $\Delta m$ said above. 

Thus very small accreted lumps do not affect our average 
$\langle L\rangle$, due to the small associated 
mass accretion rate. On the other hand, merging events involving 
 comparable partners (though contributing $\sim 1/2$ of the total mass) 
affect the overall $\langle L\rangle$ only marginally; in fact, 
such events not only are few ($<10\%$), but also they involve lumps with 
temperatures $T_1$ comparable to $T_2$, and so 
produce a compression factor $n_2/n_1\approx 1$ (eq. 9). 
 The major contribution to $\langle L\rangle$ is by far ($\sim 90 \%$) 
provided by intermediate merging lumps, which yield 
an {\it integrated} contribution $\sim 1/2$ to the mass, but 
dominate the number of events and produce large 
compressions $(n_2/n_1)^2 \propto L$. 
For such events, the isotropic hydrostatic equilibrium for the ICP 
 is physically motivated and 
robust (see \S 2.3 and \S 4), substantiating our step-by-step rendition of the 
hydrodynamics. 
\section{RESULTS} 
Here we present various results from the PE model, and 
 compare them with observations. To this aim, we shall express our results 
 in terms of the observed emission weighted temperature, which we 
denote simply by $T$. Moreover, the contribution of relevant emission lines 
to $L$ (from updates of 
Raymond \& Smith 1977) has been added to the bremmstrahlung emission 
underlying the simple scaling  in eq. 10. 

\subsection{Profiles }
Our reference cluster will have a mass $m$, and 
 DM potential $\phi(r)$ as said in Sect. 2.3. 
The density and temperature profiles are given by eq. (11), and are to 
match  the shock boundary conditions at the position $r_2 \simeq R$. 
 The key quantity is the  parameter $\beta$ defined by eq. (12); 
its average value and scatter are {\it predicted} by the PE model 
 (using convolutions analogous to eqs. 13-14)
 to be as  shown in fig. 2. 
Note that $\beta (T)$ grows slowly with the temperature;  
we obtain values ranging from about $\beta = 0.5$ at the group scale to 
$\beta \approx 0.9$ for rich clusters, see fig. 2. 
These values  imply that in our model the ICP profiles  
 are smoother and more extended than the DM's, an effect becoming 
more pronounced in downscaling from clusters to groups. 

In fig. 3  we show the baryonic fraction $f_{2}$ (integrated out to 
the shock) as a function of $T$ and for different values of $\gamma$. 
 The polytropic index $\gamma\geq 1$ describes the equation of state 
for the ICP. An upper bound to it arises if 
the overall thermal energy of the ICP is not to exceed its 
gravitational energy, with only minor contributions from 
other energy sources, like radiosource heating  or energy transfer 
from DM to ICP, as discussed in Sect. 4. 
The thermal and the gravitational energy are computed using the profiles 
in eq. (11),   and their ratio is given in fig. 4 to show that 
the {\it upper} bound  $\gamma\leq 1.3$ holds. 

In fig. 5 we show temperature profiles $T(r)$ for different values 
 of $\gamma$, in terms of the normalized coordinate $x=r/R$. 
It turns out that 
observations by Markevitch et al. (1997) are consistent with the 
$T(r)$  predicted when $\gamma = 1.2\pm 0.1$, in our allowed range. 
Hereafter we shall focus on 
 $\gamma=1.2$. 

In fig. 6 we show the density profiles $n(r)$ for two local clusters with 
different temperatures; 
the associated surface brightness $\Sigma (r)$ is shown in fig. 7 along 
with representative data. 
It can be seen that groups have 
{\it flatter} $\Sigma (r)$ than rich clusters, an outcome persisting 
when the King or the potentials by Moore at al. 1997 are used. 
%
\subsection{Correlations} 

We show first in fig. 8 the $M-T$ relation in view of its 
important role. Note that, given the mass function, 
our flattening at low temperatures translates into a steepening 
of the corresponding temperature function; such an effect has been also 
 noted by Balogh, Babul \& Patton (1998). 

The $L-T$ correlation is given by the double convolution (13), 
and likewise for $\Delta L$ after eq. (14). 
The results are shown in fig. 9 for a tilted 
CDM spectrum of perturbations in the critical universe.
For the reasons discussed in CMT98, 
the normalization has been best-fitted on the data. 

As stressed in our previous works (CMT97, CMT98), 
the correlation we predict and show in fig. 9 is {\it not} a single power-law; 
 it starts as $L\propto T^2$ for very rich clusters with high $T$, 
but bends down with decreasing $T$, 
due to the threshold effect of the preheating temperature $kT_1\approx 0.5$ keV. 
Note that our correction to 0.3 solar metallicity tends to 
increase, if anything, the luminosities and to flatten the slope at low $T$. 

A convenient fitting formula for the predicted $L-T$ 
correlation (precise to better than 10\% for $T>2\,T_1$) is as follows:
\begin{eqnarray}
L & = & a_L\,T^{2+\alpha_L}\,(\rho/\rho_o)^{1/2} \\
a_L & \propto & 
3.8\,\Omega_0^{0.3}\,(1+z)^{0.22/\Omega_0}+(1-\Omega_0)\,e^{-0.7\,(1+z)}
\nonumber 
 \\
\alpha_L & = & 
1.12\,(1+z)^{-0.2}\,e^{-0.25\,(T-T_1)/\Omega_0^{0.1}\,(1+z)^{0.5}} ~,
\nonumber 
\end{eqnarray}
where the luminosity is expressed in units of $10^{44}$ erg/s and the 
temperature in keV. 
 The $z$-dependence of $a_L$ results from 
the interplay of the following effects: 
i) the evolution of the Navarro et al. (1997) potential; ii) the abundance 
of clusters with given $T$; iii) the evolution of the progenitor probability 
 distributions in eqs. (4) and (5). 
Such effects are small, and moreover they balance out very nearly, 
leaving the basic $z$-dependence $[\rho(z)/\rho_o]^{1/2}$. In turn, the latter 
dependence goes as $(1+z)^{0.5-1}$, considering (CMT98) the evolution of 
the overdensities  in the large scale structures -- filamentary or 
sheet-like -- hosts to the clusters. 

At temperatures substantially larger than the threshold $kT_1 \simeq 0.5$ 
keV the intrinsic,  
dynamic dispersion grows  with $T$, 
but the relative $\Delta L/L$ stays nearly constant around 
$25$ $\%$ ($2\sigma$), 
as shown by fig. 9. We have deliberately chosen to keep this figure 
simple and not to include the 
conceivable spread of $T_1$ already represented by CMT98 in 
their fig. 2; 
the effect of such a spread is to widen the dispersion below 1 keV 
adding another, large component to the intrinsic dynamic variance. 

As our analytical approach allows us to span a wide range of 
cosmologies/cosmogonies, we show in fig. 10a the 
dependence  
of $\langle L\rangle$ on $\Omega_o$ at two temperatures and at the current 
epoch. It is seen that $\langle L\rangle$ increases with $\Omega_o$ increasing; 
this is because the underlying strength of the current shocks grows 
on average as the merging rate (moderately) increases on approaching the 
critical cosmology, see Lacey \& Cole (1993). A similar behaviour is 
followed by the corresponding 
dispersion $\Delta L$, see fig. 10b. 
 
Similarly, we derive a correlation with $T$ of 
the effective size of the X-ray emission. If (following 
Evrard \& Mohr 1997) we define 
$R_X$ in terms of the isophote corresponding 1.9 $10^{-3}$ ct/s arcmin$^2$ 
in the ROSAT band (consistent with our normalization for $L$), 
we find the  $\langle R_X\rangle -T$
 correlation shown in fig. 11. We find $R_X\propto T$ 
in the range of clusters, with a steepening at the group scale and a 
flattening at large temperatures. 
A corresponding fitting formula is as follows:   
\begin{eqnarray}
R_X & = & a_R\,T^{0.5+\alpha_R}\,(\rho/\rho_o)^{-1/2}  \\
a_R & = & 
0.6\,\Omega_0^{0.1}\,(1+z)^{-1.5+0.3\,(1-\Omega_0)/(1+z)}\nonumber 
 \\
\alpha_R & = & 
0.5\,\Omega_0^{-0.6}\,(1+z)^{0.6}\,e^{-0.37\,(T-T_1)\Omega_0^{0.2}/(1+z)^{1.4}} 
~,\nonumber 
\end{eqnarray}
where $R_X$ is expressed in Mpc and $T$ in keV. 
\section{CONCLUSIONS AND DISCUSSION}  

This paper is based on hierarchical clustering; group and 
cluster formation is envisaged 
in terms of DM potential wells evolving hierarchically, 
and engulfing outer baryons by accretion of smooth gas or by merging 
with other clumps. We consider the diffuse baryonic component 
to increase as the deepening wells 
overcome the external gas energy provided by preheating stars 
or by virialized subclumps. After a merging 
episode, the ICP in the wells falls back to 
a new, approximate hydrostatic equilibrium 
(hence the name ``punctuated equilibria'').

We have modeled this complex astrophysics using an  
{\it analytic} approach based on the standard hierarchical clustering and 
comprising three main {\it steps}: the hydrostatic 
equilibrium  for the ICP is computed for a given boundary condition; 
the latter is derived from the effects on the ICP of the 
dynamical evolution of the DM halos; the intrinsic stochastic character of 
 such evolution is accounted for on convolving 
with the statistics of the DM merging histories. 
All such aspects are treated in a fashion which is necessarily simplified; 
however, the resulting model -- whose parameters are all internally or  
physically constrained -- proves to be 
 gratifyingly {\it efficient} in explaining and predicting a variety of 
observations. We shall discuss such aspects 
in turn. 

As for the ICP {\it equilibrium}, 
we note that the condition $R/c_s < t_d$ 
 (sound crossing time shorter than the dynamical time) is weakly 
satisfied. However, this is enough to ensure ICP equilibrium during 
 the intervals (at least 2/3 of the total time, see Tormen et al. 1997) when 
the DM halos themselves are in approximate dynamical equilibrium. 
In point of fact, ever since Jones \& Forman (1984) to the recent 
Cirimele, Nesci \& Trevese (1997),   
it has been recognized that hydrostatic equilibrium provides for many 
clusters a fitting description of the averaged profiles of surface brightness 
in X-rays 
 (except for the central region when cooling flows set in).  
 Even the high-resolution observations provided by ASCA 
(Markevitch 1998) and those 
being derived from SAX data can be accounted for in terms of average profiles. 
Conspicuous hot spots do occur, but only in a minority of 
sources, and then in correlation with other signs of ongoing major dynamical 
events, as discussed next. 

This body of evidence supports the case 
that the gas stays close to hydrostatic equilibrium, or falls back to it 
after a short transient from the merging event, except for the rare major 
episodes involving a partner of comparable mass. The limits to 
 the above picture may be defined with the help of 
aimed hydrodynamical simulations. 
A quantitative account of how much and how long cluster collisions 
displace the ICP out of equilibrium can be found, e.g., in the N-body 
experiment of Roettiger et al. (1998) for the case of a merging event with a 
mass ratio of 1/2.5.  
Even for such ratio (already a rare event in the hierarchical 
clustering picture) the simulation shows that some 2 Gyr after the 
event hot spots and space variations of the luminosity 
 are reduced to under $20 \%$. 

So a sequence of hydrostatic equilibria of the ICP 
 is physically motivated for all merging events except for those involving 
comparable clumps (a mass ratio larger than $\sim 1/4$). However these 
 sum up to less than  $10 \%$ in the number; in addition, 
these events yield a shock compression factor $n_2/n_1\approx 1$ 
(see eq. 9), with an overall contribution to $\langle L\rangle $ 
less than 10 \%. This is actually the precision level of our model. 

Note that these considerations also support the use of the polytropic relation 
$T\sim n^{\gamma-1}$. 
In fact, when equilibrium holds, a macroscopic $\gamma=dln\,p/dln\,n$ 
 may be always defined in principle to describe the ICP state; 
the question concerning whether this is constant on scales $\gtrsim 0.1$ 
Mpc can be probed with observations. In fact, as we discuss later on 
in this Section, observed temperature and surface brightness profiles 
agree well with those predicted by 
the polytropic equation of state with $\gamma\approx 1.2$.
 
The hydrostatic equilibrium is described by a first order differential 
equation (see \S 2.3) requiring one condition at the {\it boundary}. 
Physically, this is provided by the place where a shock converts 
most of the kinetic and gravitational energy of the inflowing 
colder gas into thermal energy, as it must occur for the accreted ICP 
to be contained in the well.
The boundary condition may be referred to in terms of the stress balance 
$P_2= P_1 + n_1\, m_H \, v^2_1$, one conservation law contributing to eqs. 
(8) and (9). If it were somehow possible to 
shut off the r.h.s. completely, the intracluster medium 
would expand (the more so the closer is its state to isothermal, and the 
shallower is the potential at the shock position; the density would decrease 
everywhere including the centre and $L \propto n^2$ would quench considerably
over a sound crossing time. However, we shall argue next that the dynamic 
stress acts steadily.

In closer detail, the shock jump conditions are set in terms of 
$T_2/T_1$, basically 
the height of the current potential well (see eq. 8) compared with the 
 thermal energy of the infalling gas. The latter is 
initially due to stellar preheating (of nuclear origin); 
 then it is increased to the virial value (of gravitational origin) when 
the accreted gas is bound in 
DM subclumps. So the preheating sets an effective threshold 
$kT_1 \sim 0.5$ keV to gas inclusion, which 
breaks the self-similar correlation $L\propto T^2$ 
not only in its vicinity but also up to a few keV. 
In our model, this occurs through the specific dependence of $n_2/n_1$ on 
$T_2/T_1$ at the cluster boundary
holding for spherical shocks, strong or weak. 
This is a fair representation for the conditions prevailing when 
the cluster growth occurs by nearly isotropic 
accretion of smooth gas or of many small clumps, as shown 
by a sequence of spherical hydrodinamical simulations up to the recent one by  
Takizawa \& Mineshige (1997). Surely, this representation looks as   
a rather crude approximation 
 to the aftermaths (lasting up to 2 Gyr) of major merging events, 
as those simulated in detail by Roettiger et al. (1998). 
But in point of fact, our average quantities, their scatter and the profiles 
agree with the data over the whole range of 
$T$ as we stress next, while 
an explanation in point of principle is offered thereafter. 

For example, the equilibrium parameter $\beta (T)$  
is set by the boundary conditions in terms of shock strengths 
to values which increase from about 
0.5 at the group scale  to about $0.9$ for rich  clusters, see fig. 2. 
Correspondingly, the surface brightness profiles in groups --  
beyond the generally larger observational noise --  ought to be 
flatter than in rich clusters (see figs. 6, 7). 
In fact, similar values are obtained 
  from fits to the brightness $\Sigma (r)$ observed in groups 
 and in rich clusters ever since Kriss, Cioffi \& Canizares 1983 and 
Jones \& Forman 1984. More recently, a similar trend has been found 
from spectroscopic measurements of $\beta$ by 
 Edge \& Stewart (1991), by Henriksen et al. (1996),  
and by Girardi et al. (1998). 
We add that -- as another straightforward consequence of 
the threshold $T_1$ -- our model groups differ from rich clusters also  
for their lower, average baryonic fraction (see fig. 3) in accord with the 
average values  inferred by Reichart et al. (1998) from observations, 
however noisy. 

The other parameter of the ICP equilibrium, namely the polytropic index 
$\gamma$, is constrained to the 
rather narrow range $1\leq \gamma< 1.3$. 
The upper bound holds if the ICP thermal energy content is not to exceed 
its gravitational energy in the DM potential (see figs. 4); 
it may be 
extended to 1.4 only if kinetic energy transfer from residual clumps in DM 
to ICP contributes more than 20\%, the limit set by Kravtsov \& Klypin (1998).
 Values of  $\gamma$ around $1.2$ imply the entropy distribution $S(r)
\propto log[T(r)/n^{\gamma-1}(r)]$ to have a neat central minimum, 
 in accord with the notion of dominant entropy deposition by shocks 
in the outer regions 
 (see  David, Jones \& Forman 1996; Bower 1997) 
against the central contribution deposited by 
supernovae.  In fact, specific calculations based on entropy production at the 
shock show that $\gamma\approx 1.2$ holds with little variations from clusters 
to groups (Tozzi  \& Norman 1999, in preparation). 
Values of $\gamma\approx 1.2$ turn out 
 to yield  temperature gradients (see fig. 5) consistent not only with 
the results from advanced simulations of rich clusters 
(Bryan \& Norman 1998), but also with aimed recent observations 
by Markevitch et al. (1997) (see also Fusco Femiano \& Hughes 
1994). Preliminary data from SAX  (S. Molendi, private communication) 
indicate in some cases a somewhat flatter central 
 gradient, but still consistent with the range of $\gamma$ given above. 
That such gradients cannot be realistically traced back 
to imperfect thermalization of the electrons has been argued by 
Ettori \& Fabian (1998) (for a discussion see also 
Takizawa \& Mineshige 1998b). Note that the arguments may be reversed,
opening an interesting perspective to gauge the baryon thermal history 
and its link with the DM dynamics, see fig. 6; in fact, 
very steep DM cusps would require larger values of $\gamma$ to fit the 
core-like shape of $\Sigma (r)$, but this in turn would produce steep 
profiles of $T(r)$ and 
imply additional central inputs of energy and entropy, leading to  
a strongly bent $L-T$ relation. 

As for the {\it statistical} aspects governing 
average correlations and their scatter, these are derived from  
convolution  of the boundary conditions over the merging histories. 
 We have already predicted and discussed the $L-T$ correlation in our previous papers 
(CMT97; CMT98); here we only note, and illustrate in fig. 9, 
 the points discussed by 
  Markevitch  (1998), Allen \& Fabian (1998), namely, 
 that once the effect 
 of large cooling flows is removed or accounted for, 
the average correlation is flattened to a slope around 2.5 and the 
 scatter is reduced down to 13 \% ($1\sigma$), both in good agreement with the 
intrinsic, dynamic scatter from the model. 

Here instead we expand on the $R_X-T$ correlation, see  fig. 11. 
That our model predicts $R_X\propto T$ in the range of rich clusters with a 
steepening at the group temperatures, is due both to the 
 non selfsimilar form of the shock strength as a function of $T$, 
and to the shape of $\beta (T)$ discussed above. 
The results agree with the data, 
whilst all the self-similar computations (including the simulations 
without preheating discussed by Evrard \& Mohr 1997) yield 
$R_X\propto T^{0.5-0.7}$. The dispersion 
we find from the average over the merging histories also compares well with 
 the existing data.  

To understand such overall {\it effectiveness} of the 
model one has to consider two  features of the hierarchical clustering: 
i) the main contributions to the growth of cluster-sized 
DM halos is given by the many 
lesser, closely isotropic events (see fig. 1), which are described 
 well by the model; ii) such events 
contribute the most to statistics like 
the $\langle L\rangle -T$ correlation, as can be seen 
on examining the convolutions with the complete merging 
histories represented by eqs. (13) and (14). 
The argument reduced to the bones goes as follows: the average 
 $\langle L \rangle \propto \langle (n_2/n_1)^2\rangle$ may be expressed by 
summing the values $(n_2/n_1)^2$ after a merging episode, 
weighted with the probability $\Pi$ of a given mass ratio  (see fig. 1c), 
and with the mass increment $\Delta M/M$. Considering a cluster 
with $M\approx 10^{15}\,M_{\odot}$, the rare events ($\Pi\approx 0.15$) 
corresponding to $\Delta M/M\approx 0.5$ yield 
$(n_2/n_1)^2\approx 1$; instead, events with $\Delta M/M\lesssim 0.1$ 
have $\Pi\gtrsim 0.6$ and yield $(n_2/n_1)^2\gtrsim 6$ (from 
eq. 9 averaged over the merging histories), whose product 
 makes an overwhelming contribution to the average. 

In summary, our picture envisages the {\it combined} effect over the 
effective time $\Delta t$ of all shocks which overlap. Barring the lumps too 
small to overlap and yield an appreciable mass accretion rate, and those 
too warm and too few to yield relevant compressions, 
our physical picture 
is focused onto a nearly {\it continuous} accretion of intermediate and colder 
lumps overlapping within a sound crossing time. From these, the ICM 
feels a nearly steady external pressure with only minor fluctuations. 
Such pressure exerted at the cluster boundary 
sets the internal density via 
the connection between boundary and centre provided by 
hydrostatic equilibrium, and so determines the steady $\langle L\rangle$, 
possibly varying on cosmological time scales. 

Finally, we stress the {\it efficiency} of the present analytical approach 
in making predictions for a wide range of cosmologies/cosmogonies. 
 These are easily spanned in terms of the  the 
convolutions (13) and (14) over the merging histories to yield 
the dependence of the amplitude, shape and scatter 
of the predicted correlations $L-T$, $R_X-T$ on the 
 cosmology as given in fig. 10 and fitted with eqs. 15 and 16. 
 Such histories are dominated by those merging events between very 
{\it unequal} clumps (with $T_1\ll T$) 
 which occur {\it close} to the observation time (Lacey \& Cole 1993), as 
 shown in fig. 1b.  Though the 
merging rate does depend on $\Omega_o$, the sum over time 
 of the merging events is weakly dependent on it, and so do 
the average luminosity and the dispersion (see fig. 10).  

The advances attained by the PE model are as follows. 
The free parameters (the central density $n_c$, $\beta$ and $\gamma$) 
of the previous  hydrostatic models are now computed or
constrained. The boundary conditions that yield $n_c$ and $\beta$ are 
derived from an approximate rendition of the hydrodynamics, and are related to  
the DM dynamics. The stochastic character of the latter
 implies variance in the merging histories even at given $T$, 
and this is enhanced by 
the $n^2$ dependence of the emission to yield the intrinsic 
scatter expected in 
$L$. The resulting, narrowly constrained model predicts 
temperatures declining outwards, and -- 
in scaling down from rich clusters to groups -- smaller $\beta$ and 
shallower brightness profiles, 
decreasing baryonic content on average, and the $L-T$ relation bending down 
strongly on approaching the preheating threshold $kT_1 \sim 0.5$ keV. 

Actually, any reasonable spread in such threshold 
as discussed by CMT98 implies for groups an increased luminosity dispersion 
$\Delta L/L > 25\%~ (2\sigma)$ along with a considerable 
scatter $\Delta f /f\approx \Delta T_1/T_1$ in the baryonic   
fraction, apart from the larger 
uncertainties affecting the group observations. 
We have deliberately chosen to keep the model simple and 
to implement here neither such spread nor the $z$-dependence of $T_1$ 
which is expected for $z\gtrsim 1$, corresponding to the 
star formation rates at such early  $z$. 
We plan to expand on such issues while the high-$z$ data are drawing near. 

{\it Acknowledgements}: we are indebted to M. De Simone and D. Trevese 
for communicating their data prior to publication, to F. Governato for 
discussing with us his high-resolution N-body 
simulations, and to S. Molendi for several 
informative discussions concerning the temperature profiles from SAX. 
Thanks are due to our 
referee C. Lacey for pointing a number of errors and omissions 
 in the MS, and for having stimulated us to 
 clarify the exposition of several important points. 
 Grants from ASI and from MURST are acknowledged.

\newpage

\section*{FIGURE CAPTIONS}
\figcaption[]{
Top panel: 
a Monte Carlo realization illustrates the merging history of a DM halo 
 with final mass $10^{15}\,M_{\odot}$. 
\hfill\break
Middle panel: 
for the same halo it is shown 
the probability distribution of progenitors with mass $M'$ at the redshifts 
$z=0.03$ and $z=0.5$ that end up in a mass $M$ at $z=0$. 
The circles ($z=0.03$) and the triangles ($z=0.5$) 
represent the results from the Monte Carlo simulations, while the lines 
show the analytical results from eq. 4.\hfill\break
Bottom panel: the fraction of merging events during the last 2 Gyrs 
involving a mass $M$  and a partner with mass 
$3\,10^{13}\,M_{\odot}<M'<M/10$ (solid line); 
the lower limit arises from requiring 
 a virial temperature $T_v>0.5$ keV for the partner (see \S 2.2). 
The  dashed line is the corresponding fraction for events with $M'>M/2.5$. 
Tilted, COBE-normalized spectrum of perturbations, 
as given in White et al. (1996) in the critical 
universe with $H_o$=50 km/s Mpc. 
\label{fig1}}

\figcaption[]{
The predicted dependence of $\beta$ 
on the (emission-weighted) temperature $T$, see eq. 12 and 8. 
 DM potential as given by Navarro et al (1997), computed for 
the cosmological parameters and the perturbation power spectrum used in fig. 1.
The shaded region indicates the 2$-\sigma$ scatter due to the merging 
 histories. 
\label{fig2}}

\figcaption[]{
The predicted ratio of the baryonic fraction $f_2$ at the cluster boundary 
to the external value $f_{u}$ (White et al. 1993; White \& Fabian 1995)
as a function of the 
 temperature $T$ for $\gamma=1$ (solid line), $\gamma=1.1$ (dashed) and 
 $\gamma=1.3$ (dotted). 
\label{fig3}}

\figcaption[]{
The ratio of the thermal ICP energy $E_{ther}=3\,\int d^3r\,n(r)\,T(r)$ 
 to the ICP gravitational energy $E_{grav}=G\,m_H\,\int d^3r\,n(r)\,M(<r)/r$ 
is shown  as a function of $\gamma$ for the 
 DM potential of fig. 2 (solid line), and for the King form (dashed line) with 
 core radius $R/10$. 
The profiles $n(r)$ and $T(r)$ are provided by eq. (11). 
\label{fig4}}

\figcaption[]{
Temperature profiles for the model cluster of $10^{15}\,M_{\odot}$ at $z=0$ in 
polytropic equilibrium; $\gamma=1$ (solid line), 1.2 (dashed) and 1.66 
(dotted), see eq. (11). DM potential as in fig. 2. 
The profile is smoothed out with a filter width of 100 kpc.  
The dashed area taken from Markevitch et al. (1997) summarizes the 
observations of 30 clusters. 
\label{fig5}}

\figcaption[]{
Upper panel: The predicted ICP density profile $n(r)$ 
for a rich clusters with $kT=8$ keV (upper line),  
 and for a poor cluster with $kT=2$ keV (lower line). Radii are 
normalized to the virial radius $R$. The profiles 
are computed using the DM potential as in fig. 2, and for the ICP 
the polytropic 
index $\gamma =1.2$. 
Bottom panel: to illustrate the variations produced by 
the use of the King potential as in fig. 4, with 
$\gamma =1.1$ (dotted line); and of the Moore et al. 1997 potential,
with $\gamma =1.3$ (dashed line).
\label{fig6}}

\figcaption[]{
The surface brightness profile $\Sigma (r)$ 
from the PE model is compared with the 
 data for two oppositely extreme clusters. Top panel: A539, rather sparse and with low T, 
at $z=0.026$  (David et al. 1996, De Simone private communication).
Bottom panel: A2390, relaxed and hot, at $z=0.23$ (B\"ohringer et al. 1998). 
No attempt has been made at excluding emissions from cooling flows.  
\label{fig7}}

\figcaption[]{The predicted mass-
temperature relation for $T_1=$ 0.8, 0.5, 0.3, 0 keV, 
 from top to bottom. 
\label{fig8}}

\figcaption[]{
The average L-T correlation with its 
$2\sigma$ dispersion (shaded region)
 is shown for the PE model with the tilted CDM cosmogony of fig. 1. 
 The value $H_0=50$ km/s Mpc is assumed for the Hubble constant.  
The luminosities are 
corrected to 0.3 solar metallicity using the standard Raymond Smith 
code. Group data from Ponman et al. (1996, solid squares); cluster data from 
  Markevitch (1998, open triangles). Here $T_1 = 0.5$ keV 
with no dispersion; the effect of the latter is shown by fig. 2 of CMT98. 
\label{fig9}}

\figcaption[]{
Left panel: 
the dependence of the average luminosity 
$L\propto\langle (n_2/n_1)^2\rangle$ on $\Omega_o$. 
\hfill\break
Right panel: 
the dependence of the dispersion $\Delta L$ on  
$\Omega_o$. In both panels CDM cosmogonies are assumed, see 
 Liddle et al. (1996). 
\label{fig10}}

\figcaption[]{
The correlation of the radius $R_X$ (see text) with $T$ (emission 
weighted) is plotted for $\Omega=1$. 
 The $2\sigma$ dispersion is shown by the shaded region; 
the cosmological parameters and the perturbation power spectrum are as in 
 fig. 2. The data are from Mohr \& Evrard (1997). 
\label{fig11}}

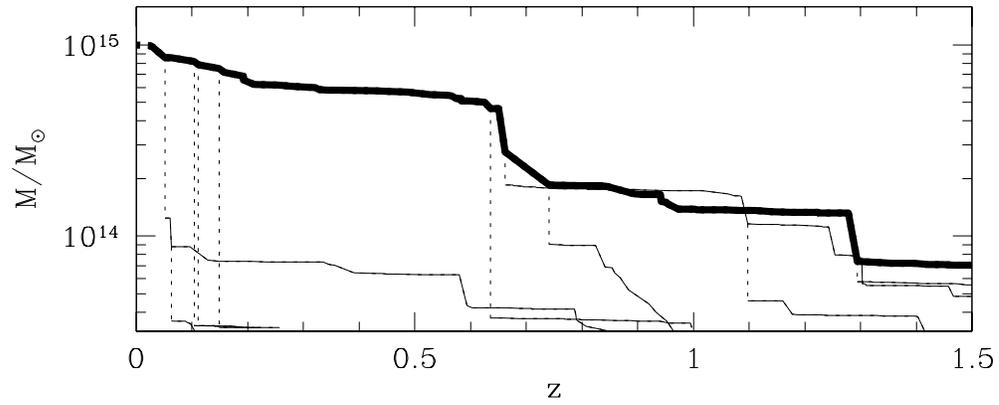
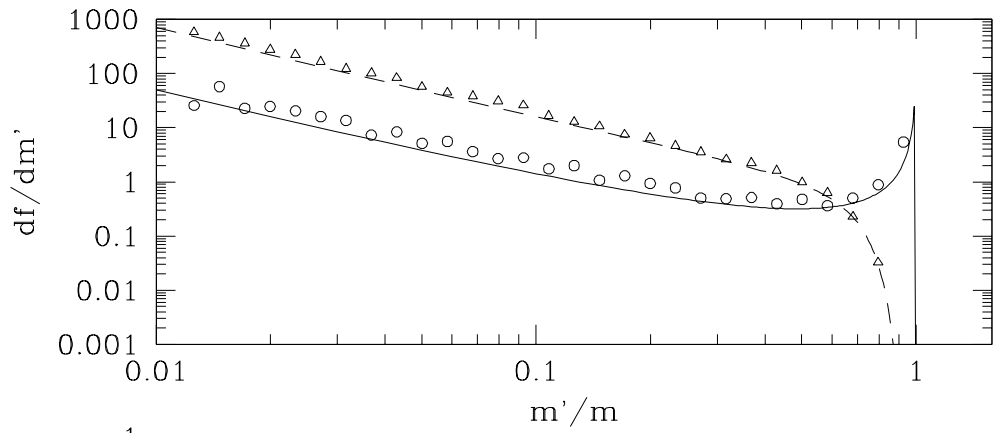
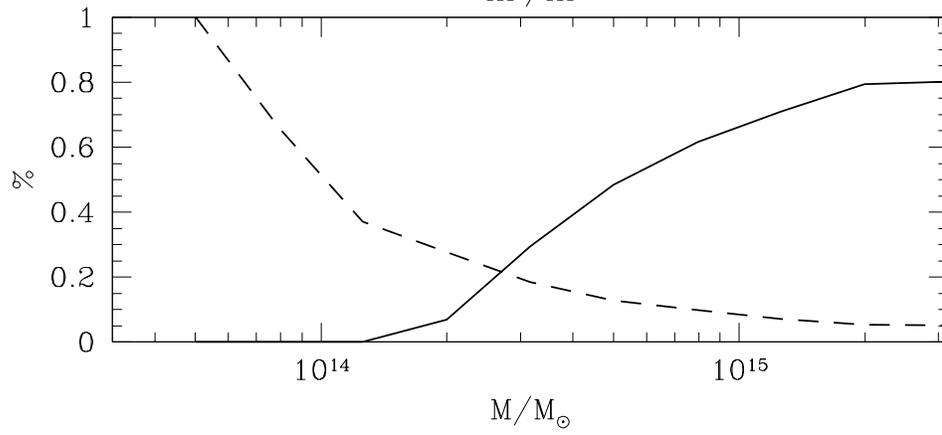
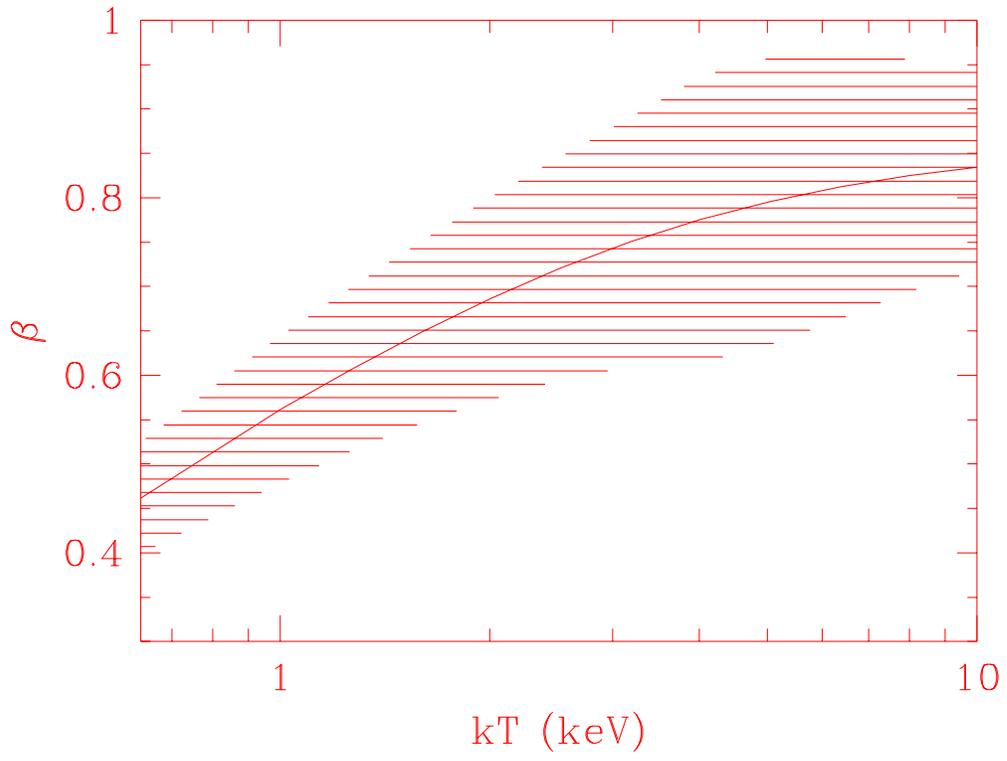
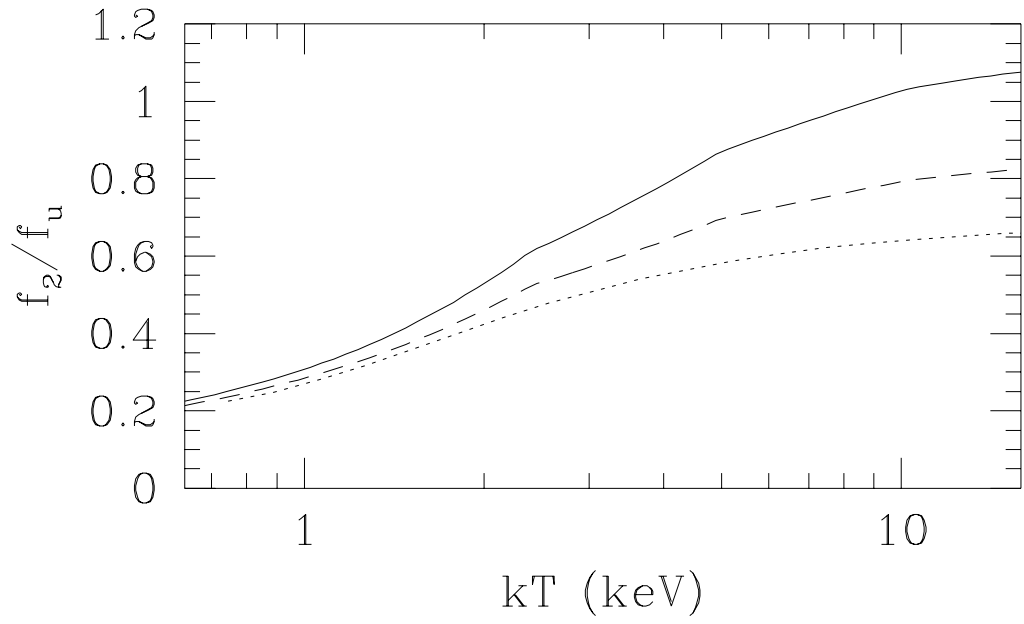
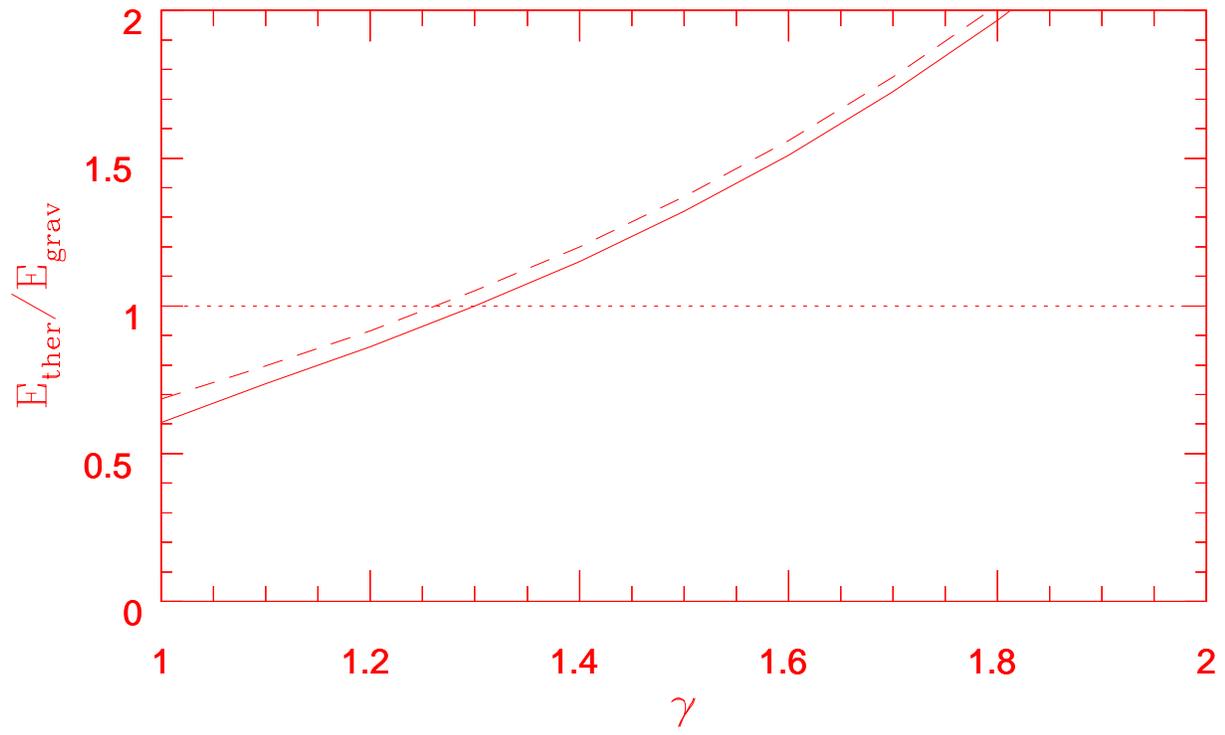
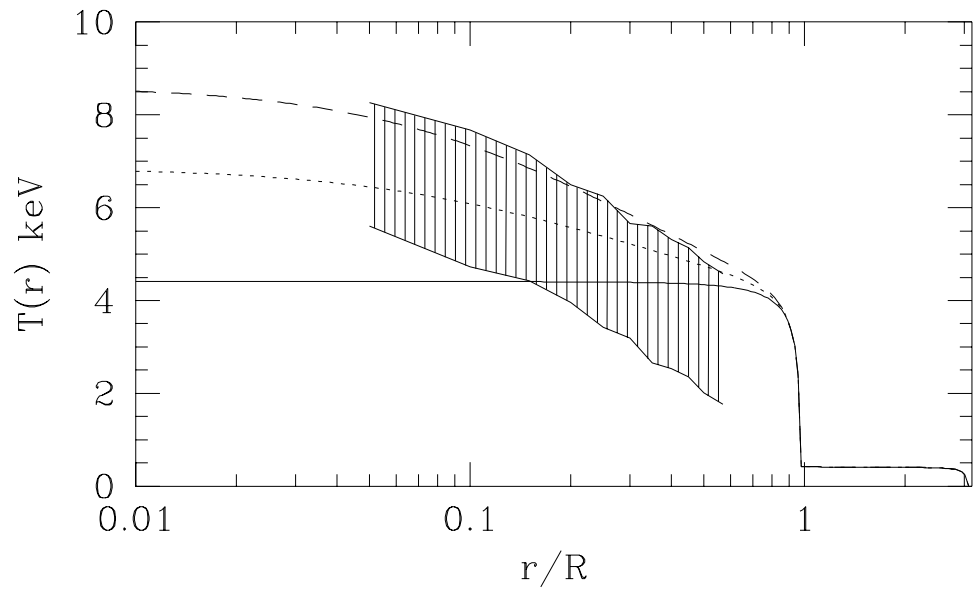
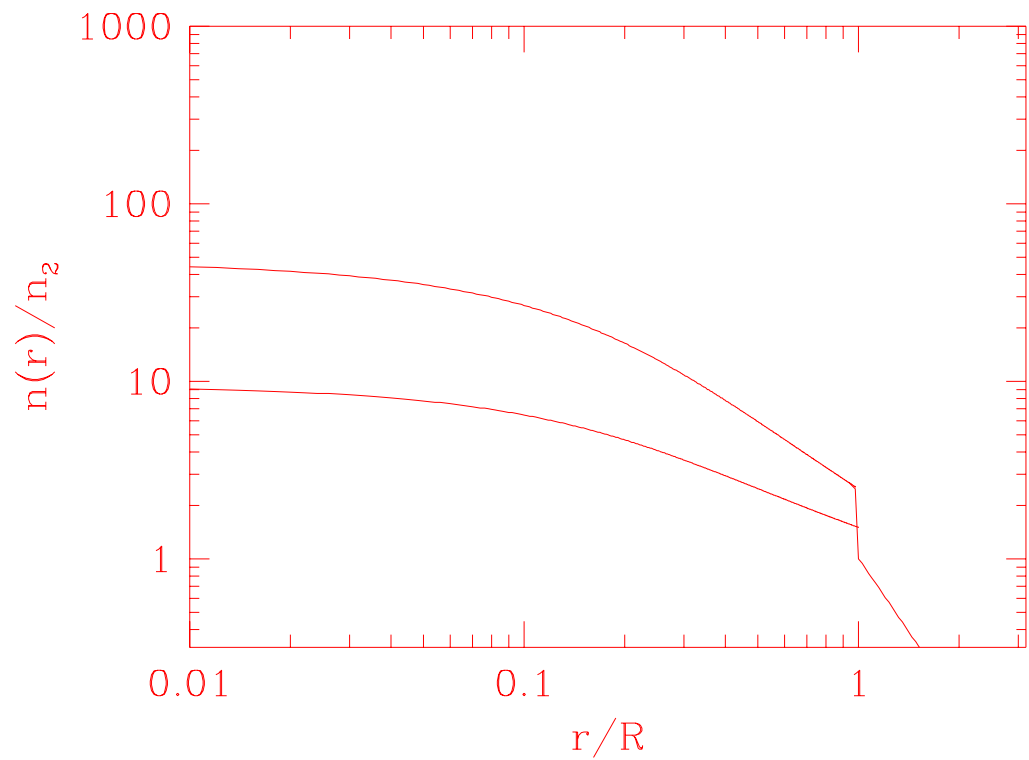
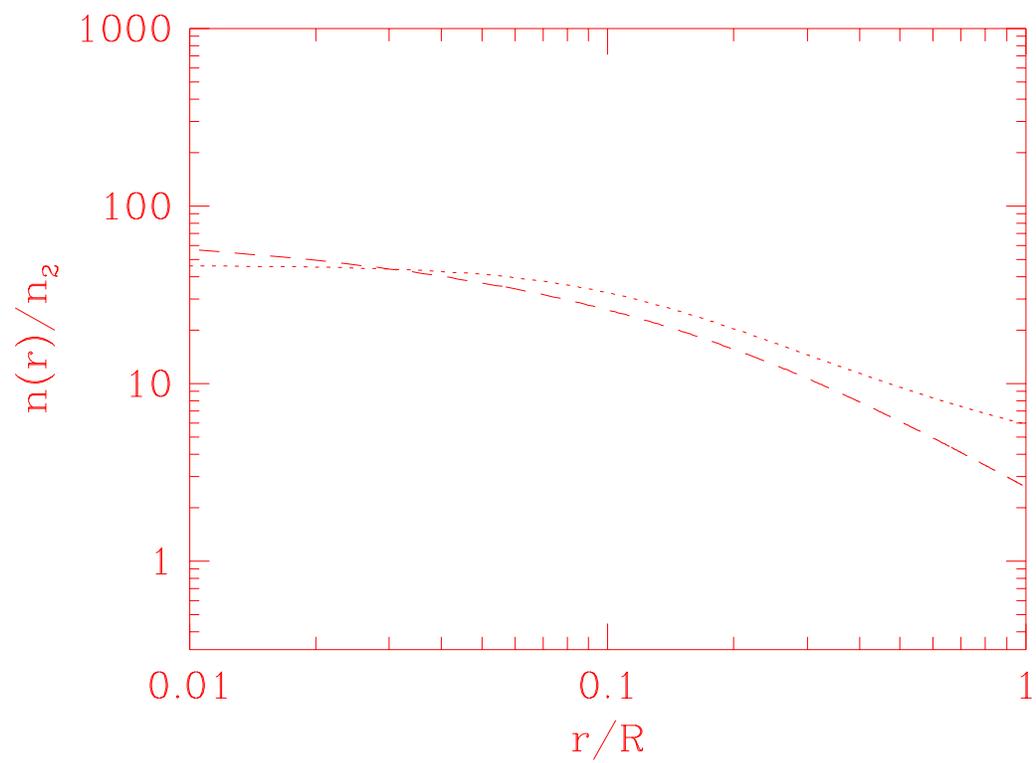
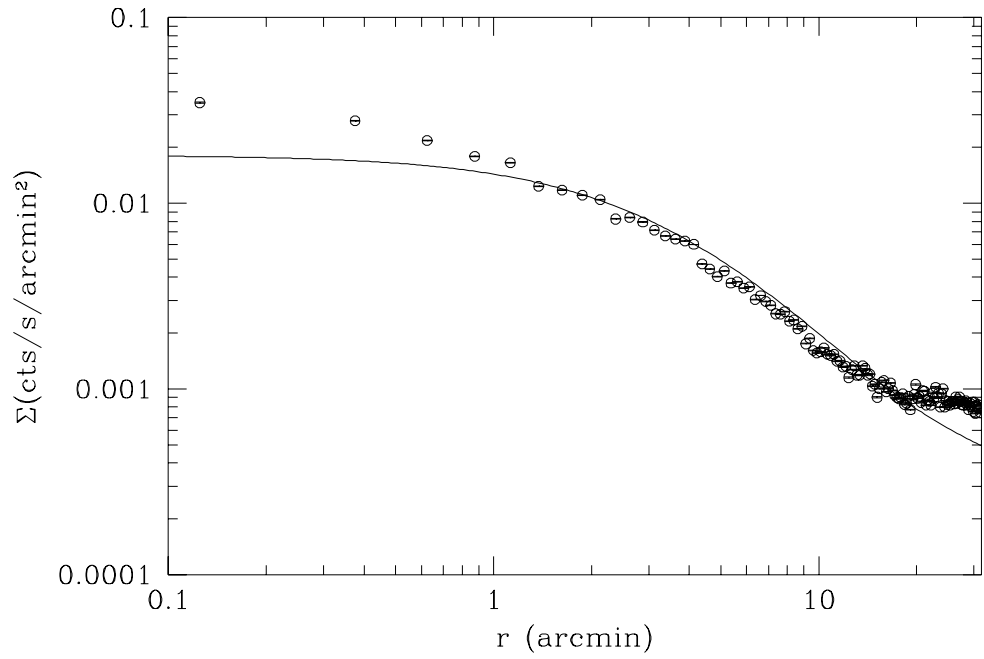
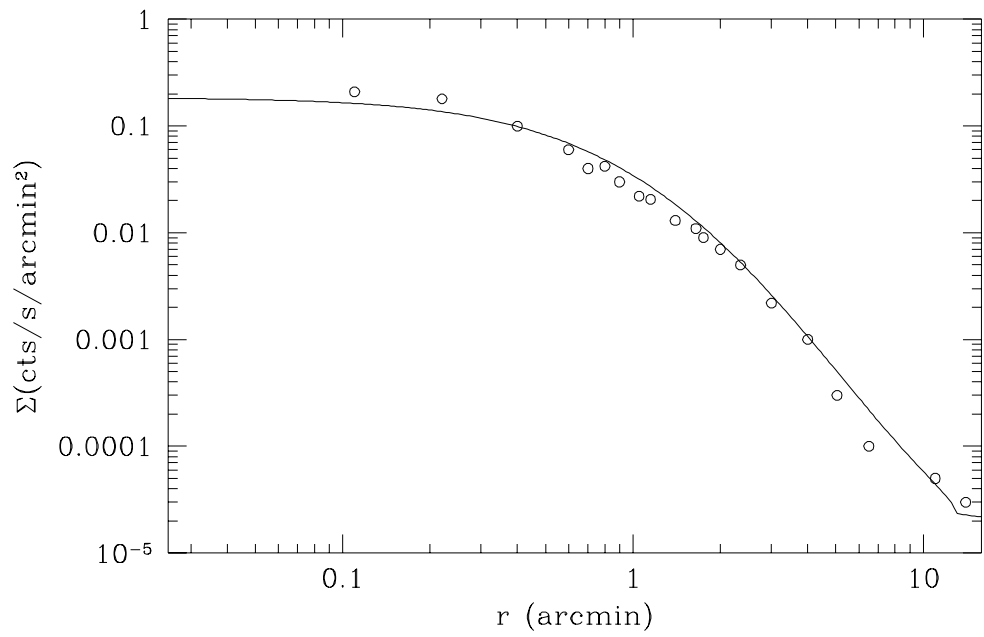
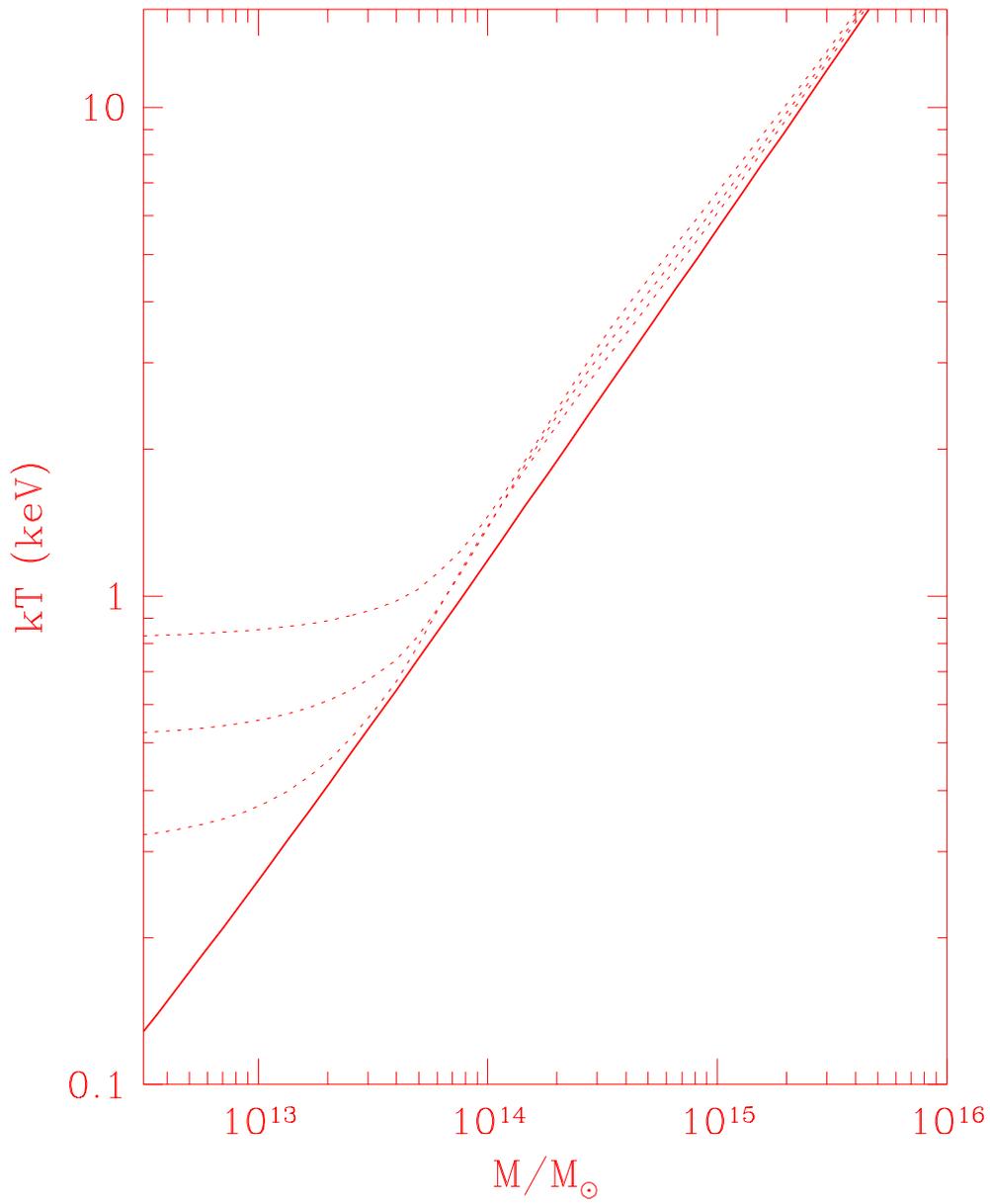
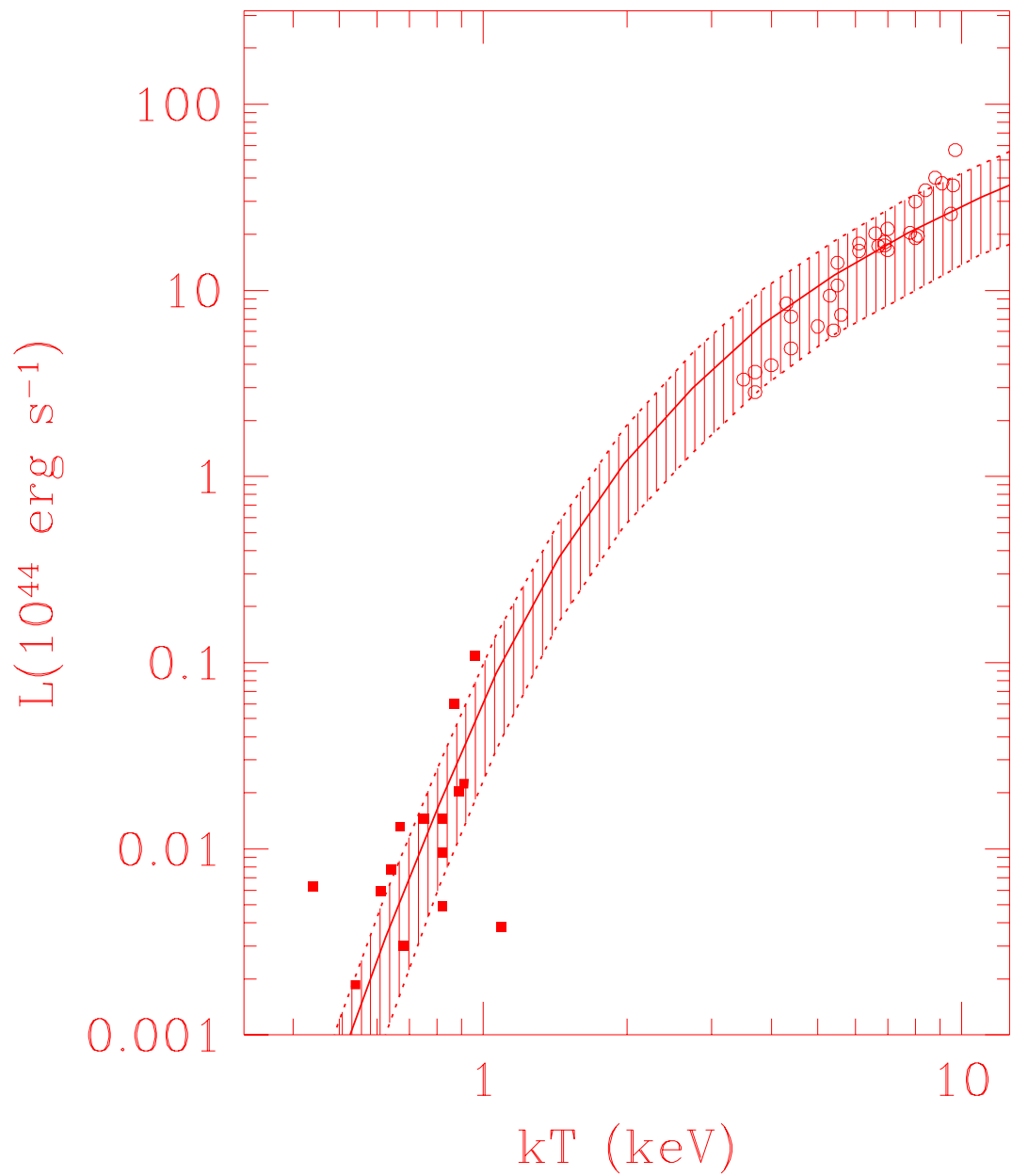
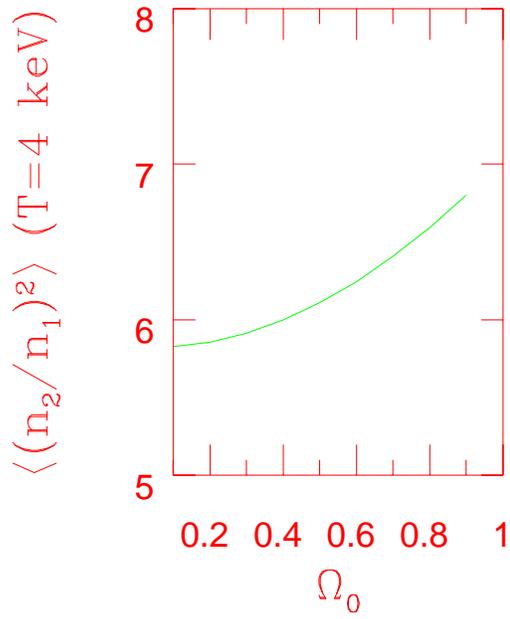
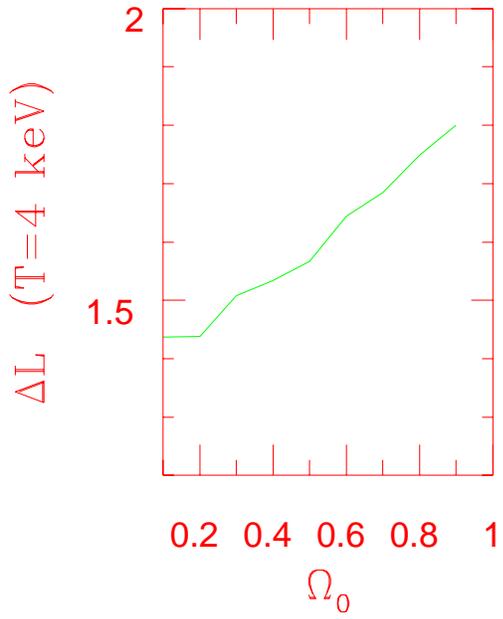
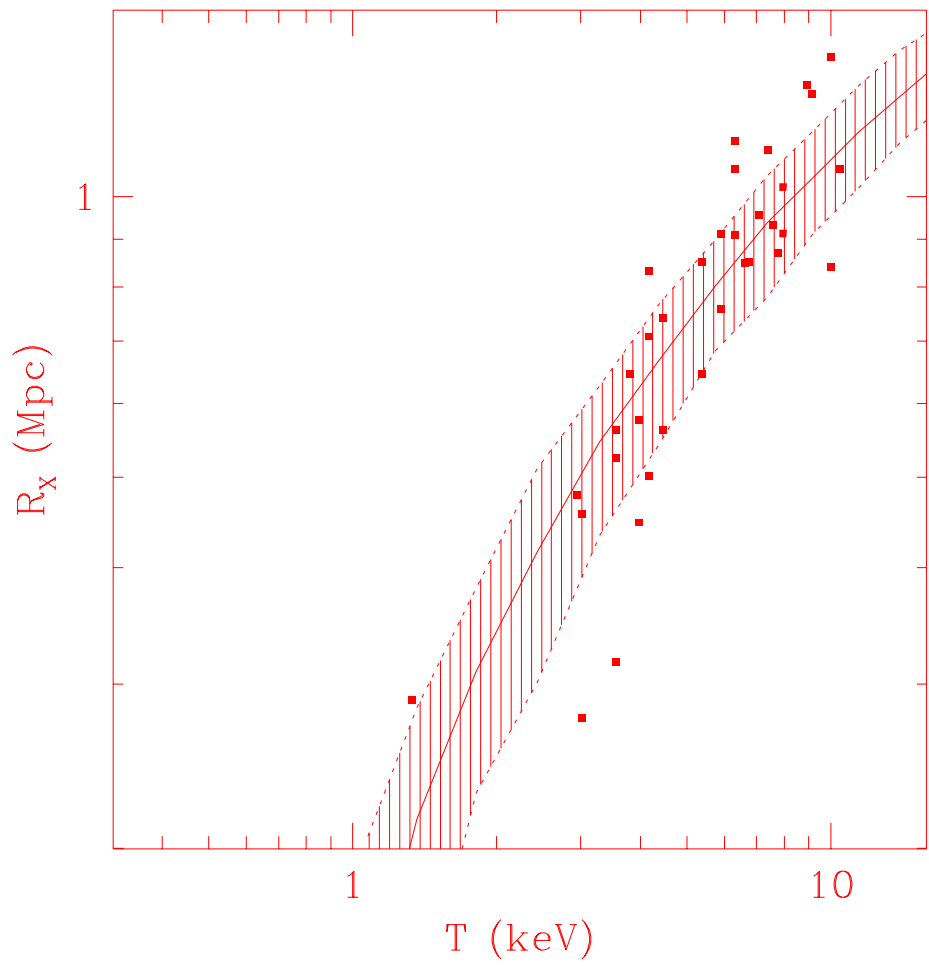
\end{document}